\documentclass[]{article}
\usepackage{graphicx}
\usepackage{amssymb}
\usepackage{a4wide}

\begin{document}


\markboth{A. Rajantie}{Contemporary Physics}

\title{Introduction to Magnetic Monopoles}

\author{Arttu Rajantie\thanks{Email: a.rajantie@imperial.ac.uk}\\
\em{Department of Physics\ Imperial College London, London SW7 2AZ, UK}}
\date{13 April 2012}
\maketitle

\begin{abstract}
One of the most basic properties of magnetism is that a magnet always has two poles, north and south, which cannot be separated into isolated poles, i.e., magnetic monopoles. However, there are strong theoretical arguments why magnetic monopoles should exist. In spite of extensive searches they have not been found, but they have nevertheless played a central role in our understanding of physics at the most fundamental level.
\end{abstract}

\section{History of Magnetic Monopoles}
Magnetism has a very long history (for more detailed discussion, see Refs.~\cite{history1,history2}), but for most of that time it was seen as something mysterious. Even today, we still do not fully understand one very elementary property of magnets, which we all learn very early on in school: Why does a magnet always have two poles, north and south? Or, in other words, why cannot magnetic field lines end? Electric field lines end in electric charges, but it appears that there are no magnetic charges. 
The purpose of this paper is to give a brief summary of the physics of magnetic charges, or magnetic monopoles, and reasons why they may exist.

The earliest description of magnetism is attributed to
Thales of Miletus, northern Greece, who reported that pieces of rock from Magnesia (naturally magnetic form of magnetite known as lodestone) 
had strange properties.  
He had also noted that amber could attract light objects such as feathers or hair after it had been rubbed, which is now understood as an effect of static electricity.
In a somewhat similar way, lodestone could attract iron, but no other material, even other metals.
Even more mysteriously, when hung from a piece of string, a piece of lodestone would always orient itself in the same way.
Unlike amber, lodestone did not need rubbing or any other preparation to acquire these properties, but when an iron needle was rubbed with lodestone, these properties were transferred to it.

Lodestone was also known elsewhere, for example in China. While the Greeks thought that the rocks were attracted to the place where they had been mined, the Chinese noticed that a piece of lodestone actually always pointed north or south. At the start of the last millennium, they already knew how to use this effect as a navigational tool in the form of a compass, and by 1187 this invention has found its way to Europe.

The attempts to explain and understand these powers of magnets were based on pure philosophical speculation until 1269 when Petrus Peregrinus of Maricourt
wrote a letter called {\it Epistola de magnete} which described experiments he had carried out on magnets and the conclusions he had drawn from them. He showed that a magnet has two opposite poles (and introduced the term ``polus'' to describe them), and that when a magnet is cut in two halves, each of them still has two opposite poles. He found that two like poles repel each other, but he failed to attribute the north-south orientation of the compass needle to the Earth's magnetism. Instead, he accepted the prevailing view that it was due to its attraction to celestial poles.

It took three more centuries before further progress was made. 
It had been assumed that whatever force makes the compass needle turn north, it acts in the horizontal direction, but
in 1576 Robert Norman found that in fact the needle tends to tilt downwards. At first he thought that he had simply failed to balance 
his needle properly, but further experiments confirmed his finding. A needle that was perfectly balanced would, after being magnetised, turn downwards. This was a clear hint 
that the force causing the orientation of the needle is due to the Earth rather than the heavens, although Norman failed to draw this conclusion from his findings. 
More importantly for our purposes, he found that the there actually no overall attraction at all, only a change in orientation. This is a direct sign that there are no magnetic charges because they would experience a force in the direction of the magnetic field.

Relatively shortly after this, in 1600, William Gilbert published a book titled {\it De Magnete}~\cite{deMagnete} in which he tried to construct the first consistent theory of magnetism, based on careful and systematic experiments. In particular, he was the first to understand that the that the orientation of the compass needle was caused by the magnetism of the Earth. This book is regarded as the start of the scientific study of magnetism.

During the following centuries, many aspects of magnetism and its connection with electricity were discovered, and theories describing it became more specific and more quantitative. 
Nevertheless, it was widely believed that magnetism was caused by two oppositely charged magnetic fluids, which consist of magnetic molecules, or magnetic monopoles as we would call them.
This idea persisted well into to the 19th century, until Andre-Marie Amp\`ere showed that magnetic fields are generated by electric currents 
and eventually Michael Faraday demonstrated that the magnetic fluids do not exist~\cite{Faraday1851}. 

Building on Faraday's findings and physical intuition about electric and magnetic fields, James Clerk Maxwell developed his theory of electrodynamics, which he published in 1864~\cite{Maxwell}. This theory was a massive breakthrough in physics, because it not only unified magnetism and electricity in one simple theory but also explained the properties of light, which it showed to be an electromagnetic wave travelling through space. Furthermore, the theory showed that the speed of light had to be constant, and therefore paved the way for theory of relativity.

In his theory, Maxwell assumed that no magnetic charges exist, but the theory did not explain why and its structure would have allowed his to include them easily.
Magnetic charges were sometimes discussed to illustrate some aspect of physics or as a convenient mathematical tool, 
but they were not considered to be real, physical objects. One exception to this was
Pierre Curie~\cite{Curie}, who speculated on the possibility of free magnetic charges in 1894.

When quantum mechanics was developed at the start of the 20th century, it first appeared to be inconsistent with isolated magnetic charges. However, in 1931 Paul Dirac
showed that in fact, quantum mechanics allows certain quantised magnetic charges~\cite{Dirac:1931kp}. Furthermore, the existence of magnetic charges would explain the observed quantisation of electric charge, which until then had been a mystery. Finally, in 1974, Gerard 't~Hooft and Alexander Polyakov showed that magnetic monopoles are predicted by all Grand Unified Theories (GUTs), which aim to describe electromagnetic interactions and the weak and strong nuclear forces by a single unified theory~\cite{'tHooft:1974qc,Polyakov:1974ek}.
The same conclusion also applies generally to more modern ``theories of everything'' (TOEs) such as superstring theory, which also include gravitation~\cite{Duff:1994an}.

From a theoretical point of view, we would therefore expect magnetic monopoles to exist.
Finding one would be an incredible breakthrough for high energy physics. They derive their properties from laws of physics at very high energies, which are far out of reach of any particle accelerator experiments.
They are absolutely stable, so they would not decay to other particles, unlike most other particles physicists are hoping to find, and they interact relatively strongly through the electromagnetic field, which means that they would be easy to study experimentally. Their discovery would therefore allow physicists to design experiments to test GUTs and TOEs directly.
Unfortunately, all attempts to find them (see, e.g.,~\cite{Nakamura:2010zzi,Giacomelli:2011re}) have failed so far, but they have still played a key role in the development of particle physics and cosmology in the recent decades.

The purpose of this article is to give a brief introduction to the physics of magnetic monopoles. For a more detailed review, see, e.g.,~\cite{Milton:2006cp,Shnir}. A comprehensive and up-to-date magnetic monopole bibliography is also available~\cite{Balestra:2011ks}.

\section{Monopole Solutions}
\subsection{Monopoles in Classical Electrodynamics}
Classical electrodynamics is described elegantly by four Maxwell's equations. 
In the absence of any electric and magnetic charges, they read,\footnote{%
In SI units, these equations also involve the permittivity $\epsilon_0$ and permeability $\mu_0$ of the vacuum, but for simplicity I am using {\it natural units} in which these are both set equal to $1$, which also implies that the speed of light is equal to one, $c=1$. Furthermore, one also sets Planck's constant to one, $\hbar=1$. This has the consequence that everything can be expressed in units of energy.
The unit of energy conventionally used in particle physics is gigaelectronvolt $1~{\rm GeV}\approx 
1.6\times 10^{-10}~{\rm J}$.}
\begin{eqnarray}
\label{equ:ME10}
 \vec{\nabla}\cdot\vec{E}&=&0,\\
\label{equ:ME20}
 \vec{\nabla}\times\vec{E}&=&-\frac{\partial\vec{B}}{\partial t},\\
\label{equ:ME30}
 \vec{\nabla}\cdot\vec{B}&=&0,\\
\label{equ:ME40}
 \vec{\nabla}\times\vec{B}&=&\frac{\partial\vec{E}}{\partial t},
\end{eqnarray}
where $\vec{E}$ and $\vec{B}$ are vectors representing the electric and magnetic fields, respectively.
In these equations the symbol $\vec{\nabla}\cdot$ is known as the {\it divergence} of the vector field,
and $\vec{\nabla}\times$ is known as the {\it curl}.
If one represents the electric and magnetic fields by field lines, non-zero divergence would indicate that field lines are ending, whereas the curl describes their curvature.
Therefore, Eqs.~(\ref{equ:ME10}) and (\ref{equ:ME30})
show that the neither electric nor magnetic field lines have start or end points or, in other words, that they are closed.
Eqs.~(\ref{equ:ME20}) and (\ref{equ:ME40})
show that time dependent magnetic fields generate electric fields, and vice versa.

These equations have a beautiful symmetry known as electric-magnetic duality: If we replace the electric and magnetic fields as
\begin{equation}
\label{equ:duality}
\vec{E}\rightarrow\vec{B},\quad
\mbox{and}\quad \vec{B}\rightarrow-\vec{E},
\end{equation}
 the equations remain unchanged. This means that the electric and magnetic fields themselves behave in exactly the same way.

Of course, electric charges exist, and electric field lines can start and end at electric charges. More precisely, the electric field around the electric charge $q$ has the form
\begin{equation}
\label{equ:CoulombE}
\vec{E}(\vec{r})=q\frac{\vec{r}}{r^3},
\end{equation}
and a charge moving in an electromagnetic field at velocity experiences the Lorentz force
\begin{equation}
\label{equ:LorentzE}
\vec{F}=q\left(\vec{E}+\vec{v}\times\vec{B}\right).
\end{equation}
The duality (\ref{equ:duality}) shows that 
if magnetic charges existed, the magnetic field around a magnetic charge $g$ would be the dual of Eq.~(\ref{equ:CoulombE}),
\begin{equation}
\vec{B}(\vec{r})=g\frac{\vec{r}}{r^3},
\end{equation}
and a magnetic charge would experience the force
\begin{equation}
\vec{F}=g\left(\vec{B}-\vec{v}\times\vec{E}\right).
\end{equation}
However, because no magnetic charges have been found, the duality symmetry appears to be broken.
This means that, in fact, the only difference between electricity and magnetism is that electric charges exist but magnetic charges do not. 

This raises the question why nature has this asymmetry. From the point of view of classical electrodynamics, there is no reason why there could not be magnetic charges, and if they did, the duality symmetry would be intact.
In other words, classical electrodynamics is perfectly compatible with the notion of magnetic monopoles, and from the aesthetic point of view it is strange that magnetic monopoles do not seem to exist, because their existence would make the theory more symmetric.

\subsection{Monopoles in Quantum Theory}

In quantum theory, the question about the possible existence of magnetic monopoles becomes more complex but also more intriguing. 
This is because it turns out that in quantum mechanics, electromagnetic forces have to be described in terms of scalar and vector potentials $\phi$ and $\vec{A}$~\cite{Aharonov:1959fk}, instead of the electric and magnetic fields $\vec{E}$ and $\vec{B}$. These potentials appear in the theory as position-dependent fields, whose variations give rise to electric and magnetic fields through the relations
\begin{eqnarray}
\vec{E} &=& -\frac{\partial \vec{A}}{\partial t}-\vec{\nabla}\phi,\nonumber\\
\vec{B} &=& \vec{\nabla}\times\vec{A}.
\label{equ:potentials}
\end{eqnarray}
In classical electrodynamics these potentials are widely used as convenient calculation tools, and 
the scalar potential is, in fact, simply the familiar electric potential which we learn about in school, but their use is optional and one can perfectly well do calculations in terms of the electric and magnetic fields, too.

The introduction of the scalar and vector potentials through Eq.~(\ref{equ:potentials}) seems to break the duality symmetry, but one should note that the potentials themselves are not physical, observable quantities. In fact there are an infinite number of different potentials that give rise to the same electric and magnetic fields. Changing from one of these physically equivalent potential configurations to another is known as a {\it gauge transformation}, and because physical quantities do not change under such gauge transformations we say that the theory has a {\it gauge symmetry}. The gauge symmetry of Maxwell's equations is known mathematically as U(1). Similar, but somewhat more complex gauge symmetries plays a central role in our current theories of particle physics. They are seen as the fundamental principle that determines the nature of the elementary particle interactions, and we will come back to them shortly.

Nevertheless, the potentials (\ref{equ:potentials}) appear to forbid magnetic charges. This is because one of the basic results in vector analysis is that the divergence of the curl of a vector field always vanishes, so that one finds
\begin{equation}
\vec{\nabla}\cdot\vec{B}=\vec{\nabla}\cdot\left(\vec{\nabla}\times\vec{A}\right)=0.
\end{equation}
Therefore, if the magnetic field is represented by the vector potential through Eq.~(\ref{equ:potentials}), then its field lines can never have end points. This appears to show that one cannot describe magnetic monopoles using the vector potential.

\begin{figure}
\begin{center}
\begin{tabular}{ll}
(a)&(b)\cr
\includegraphics[width=7cm]{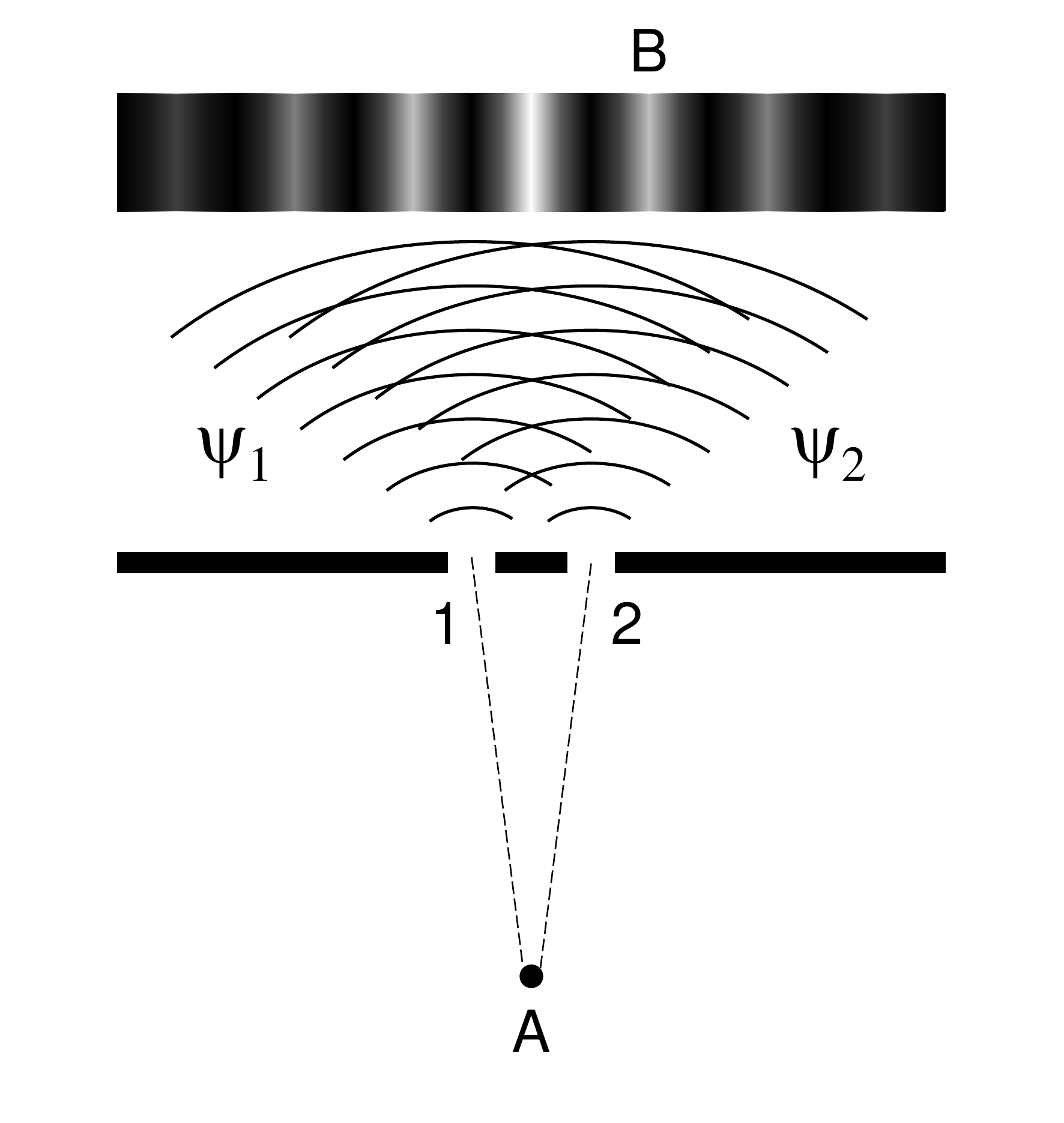}&
\includegraphics[width=7cm]{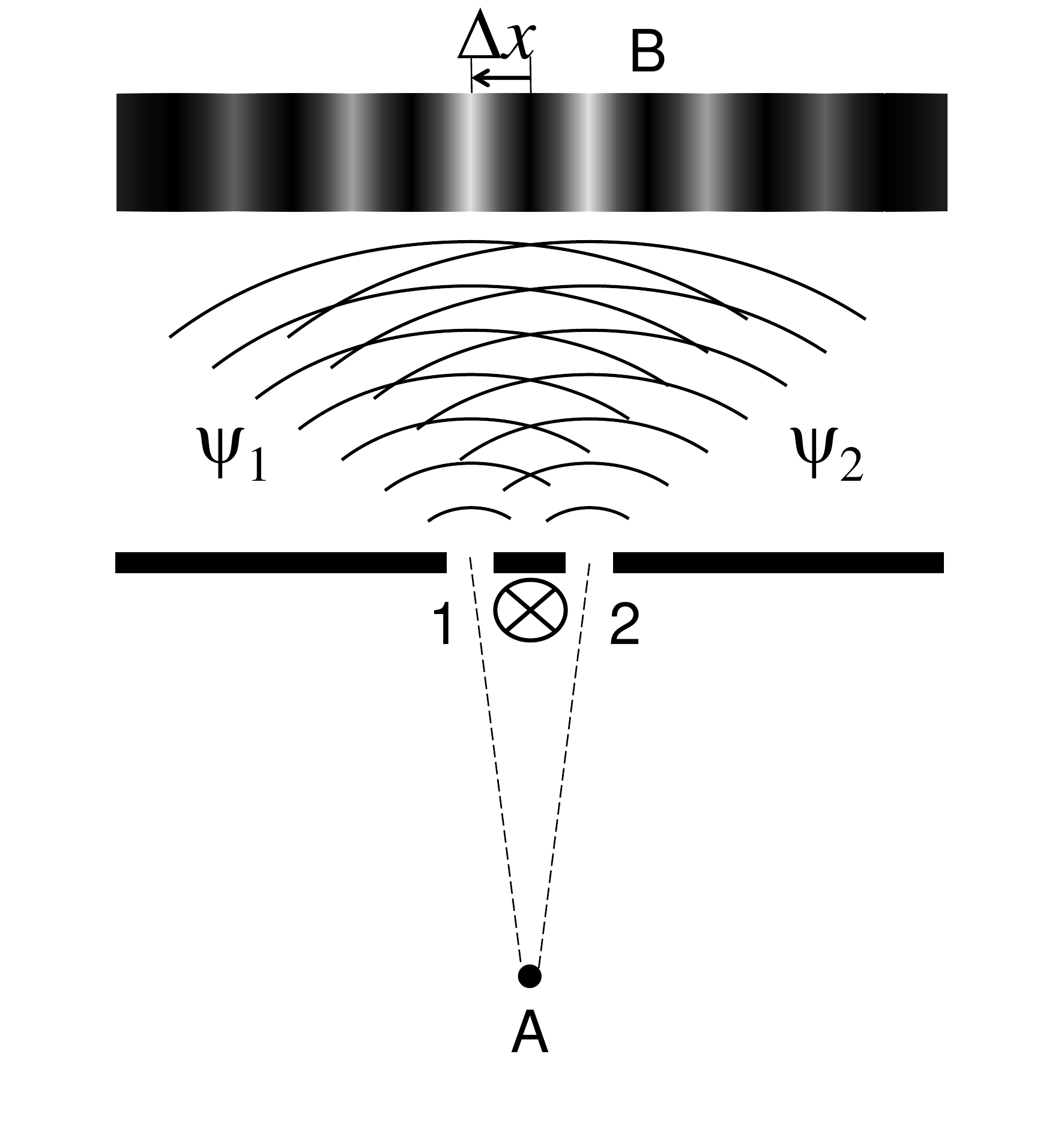}
\end{tabular}
\end{center}
\caption{\label{fig:doubleslit}
(a) In the double-slit experiment, charged particles from the source A travel through one of two slits 1 and 2, and are detected on a screen B behind them. The wave functions $\psi_1$ and $\psi_2$ interfere depending on their complex phases, producing a pattern consisting of peaks and troughs on the screen.
(b) When a solenoid $\otimes$ (or a Dirac string) carrying magnetic flux is placed between the slits, it changes the relative complex phases of the two wave functions $\psi_1$ and $\psi_2$. This shifts the interference pattern by a distance $\Delta x$ which is proportional to the flux $\Phi$.
For $\Phi=2\pi/q$, this distance is equal to the distance between peaks, and therefore the pattern is unchanged and the Dirac string is unobservable.}
\end{figure}

This restriction becomes important in quantum mechanics. Remember that in quantum mechanics a particle is described in terms of its wave function $\psi$, which is a complex number. In general, the particle does not have a well-defined position, but the probability of finding the particle at a particular position is given by the square of the absolute value of the wave function, $|\psi|^2$. The complex nature of the wave function becomes apparent in interference experiments.
For example, in the famous {\it double-slit experiment} (see Fig.~\ref{fig:doubleslit}), particles can travel through either of two parallel slits, and the final wave function $\psi(x)$ on the screen behind them is a linear superposition of the wave functions $\psi_1(x)$ and $\psi_2(x)$ corresponding to each of the two routes, $\psi(x)=\psi_1(x)+\psi_2(x)$. At places where the complex phases of the two wave function coincide, one has {\it constructive interference} and the probability is enhanced, and at places where the complex phases are opposite, the two wave function cancel each other leading to {\it destructive interference} and reduced probability.

When one describes as electrically charged particle in quantum mechanics, the complex phase of the wave function depends on the vector potential $\vec{A}$. More precisely, when moving in space, the complex phase changes by the rate given by the component of the vector potential parallel to the motion. This is compatible with the gauge symmetry, because neither the complex phase nor the vector potential are directly observable quantities. Nevertheless, this shows that 
one needs to use the vector potential instead of the magnetic field in quantum mechanics.
Because the vector potential cannot describe a magnetic monopole, does this mean that quantum mechanics prohibits the existence of magnetic charges?

In 1931, Paul Dirac showed that it does not~\cite{Dirac:1931kp}. First, he was able to find a vector potential that describes a magnetic monopole. He did this in essentially the same way Faraday had constructed a ``magnetic monopole'' for his experiments 110 years earlier~\cite{Faraday1844}: In order to investigate the motion of a magnetic monopole in the presence of an electric current, Faraday used a long magnet in a vessel filled with mercury, in such a way that one of the poles was under the surface and the other above it. That way the upper pole behaved essentially like a magnetic monopole.

To be more precise, let us consider a long and thin solenoid, i.e., a long straight coil of wire. When an electric current runs through the solenoid, it produces a magnetic field inside it. 
Because the field lines cannot end, this field spreads out in all directions from the end of the solenoid. 
It is useful to describe this in terms of the magnetic flux $\Phi$, which is the magnetic field integrated over a cross-sectional surface. Because field lines cannot end, the flux coming out of the end of the solenoid has to be equal to the flux $\Phi$ inside it.

If the length of the solenoid is much greater than its width, the shape of the magnetic field around the end of the solenoid looks exactly like that of a magnetic monopole with magnetic charge $g$ equal to the magnetic flux, so that $g=\Phi$. Along the length of the solenoid, the magnetic field is confined inside it, apart from the monopole fields at the two ends. 
Because there are no magnetic charges in this system, one can write down the vector potential for it without any difficulty.
Mathematically, one can imagine first making the solenoid infinitesimally thin and long so that the other endpoint is infinitely far away, and then forget about the solenoid itself and only keep the vector potential, which can be written as
\begin{equation} \vec{A}(\vec{r})=\frac{g}{4\pi|\vec{r}|}\frac{\vec{r}\times\hat{k}}{|\vec{r}|-\vec{r}\cdot\hat{k}}.
\end{equation}
Here $\hat{k}$ is a unit vector pointing in the direction of the solenoid.
This vector potential describes a magnetic monopole connected to an infinitesimally thin line (known as a {\it Dirac string}) carrying the magnetic flux to it.
Because in classical electrodynamics magnetic fields only affect particles that are inside the field, the Dirac string is unobservable if its thickness is zero, and therefore the vector potential really describes an isolated magnetic monopole. 
The price one has to pay is that the vector potential is singular along the Dirac string.

However, the situation is more complicated in quantum mechanics. As we saw earlier, the vector potential $\vec{A}$ affects the complex phase of the particle's wave function, and therefore the Dirac string can give rise to an interference effect even if the particles are not actually touching it. If we consider again the double slit experiment, and imagine having a Dirac string placed between the two slits, the wave functions corresponding to the two slits would pick up a different complex phase factor because they are travelling through a different vector potential. This would shift the interference pattern on the screen left or right depending on the magnetic flux~\cite{Aharonov:1959fk}.

It is fairly easy to calculate the extra complex phase $\Delta\theta$ a particle picks up. It is simply given by integrating the vector potential $\vec{A}$ along the trajectory and multiplying this by the electric charge $q$ of the particle. We can write this as
\begin{equation}
 \Delta\theta_X=q\int_{A\rightarrow X\rightarrow B} d\vec{r}\cdot\vec{A},
\end{equation}
where $X=1$ or $2$, depending on which slit the the particle goes through.
The interference pattern on the screen will only be affected by the Dirac string if the particles passing through different slits pick up a different complex phase. We can write this difference as
\begin{equation}
 \Delta\theta = \Delta\theta_1-\Delta\theta_2=q\oint_{A\rightarrow 1\rightarrow B\rightarrow 2\rightarrow A}
d\vec{r}\cdot\vec{A},
\end{equation}
which represents and integral along a closed paths from $A$ to $1$ to $B$ to $2$ back to $A$.
A general result from vector analysis known as Stokes's theorem shows that the integral is equal to the magnetic flux $\Phi$ through the closed path. Combining this with the observation above that the flux is equal to the magnetic charge $g$ of the monopole, we find that the complex phase difference between the two slits $\Delta\theta$ is related to the charge of the magnetic monopole by the simple relation
\begin{equation}
 \Delta\theta=qg.
\end{equation}
The Dirac string of any magnetic monopole with non-zero magnetic charge $g$ would therefore generate a non-zero difference in the complex phase. Now, it is important to realise that the complex phase is only defined modulo $2\pi$. Two complex numbers whose phases differ by an integer multiple of $2\pi$ are equal. Therefore, the Dirac string is only observable if the phase difference $\Delta\theta$ is not an integer multiple of $2\pi$. Conversely, if the electric and the magnetic charges $q$ and $g$ satisfy~\cite{Dirac:1931kp}
\begin{equation}
 \frac{qg}{2\pi}\in\mathbb{Z},
\label{equ:Diracquant}
\end{equation}
then the Dirac string is unobservable and the electrically charged particle only feels the magnetic field of the monopole.
This is known as the {\it Dirac quantisation condition}.

If there are several different types of electrically charged particles, then in order for the Dirac string to be completely unobservable,  the electric charge of each of the particle species has to satisfy Eq.~(\ref{equ:Diracquant}).
Dirac realised that this is only possible if the electric charge is quantised, meaning that the electric charge of any particle is an integer multiple of some elementary charge, which itself satisfies Eq.~(\ref{equ:Diracquant}). 
This is a remarkable finding because electric charges are indeed quantised in nature: They are all integer multiples of the electron charge $e$ (or one third of it, $e/3$, if one includes quarks but they are not observed as individual particles). Therefore, Dirac had found that this could be explained if magnetic monopoles exist, and, perhaps emboldened by his earlier successful prediction of the positron, he concluded that it would be surprising if magnetic monopoles did not exist~\cite{Dirac:1931kp,Dirac:1948um}. 

Even though we now have a much better understanding of elementary particles and their properties that in Dirac's time, and even though there are other ways one can try understand the quantisation of electric charge, they all seem to predict magnetic monopoles at the same time, so it seems that Dirac's insight may well have been valid.
For example, string theorist Joseph Polchinski~\cite{Polchinski:2003bq} has argued that any theory that explains charge quantisation would inevitably predict the existence of magnetic monopoles.

Although, as Dirac showed, magnetic monopoles can be consistently described in quantum theory, they do not appear automatically unlike the positron. Quite the contrary, it is actually difficult to add magnetic monopoles to quantum electrodynamics, the quantum version of Maxwell's theory~\cite{Schwinger:1966nj}. This can be understood as the flip side of Dirac's quantisation argument: In quantum electrodynamics electric charges are not naturally quantised, and therefore including magnetic monopoles has to change the structure of the theory significantly to force the quantisation condition (\ref{equ:Diracquant}).

\subsection{'t~Hooft-Polyakov monopoles}
\label{sec:tpmono}
By 1970s, physicists had moved on from quantum electrodynamics to develop a theoretical description of strong and weak nuclear forces, which would eventually become our current Standard Model of particle physics. The guiding principle was the concept of gauge symmetry that we encountered in the context of Maxwell's theory of electrodynamics, and although the gauge transformations involved are more complicated, the structure of these theories is very similar to Maxwell's theory. 

\begin{figure}
\begin{center}
 \includegraphics[width=6cm]{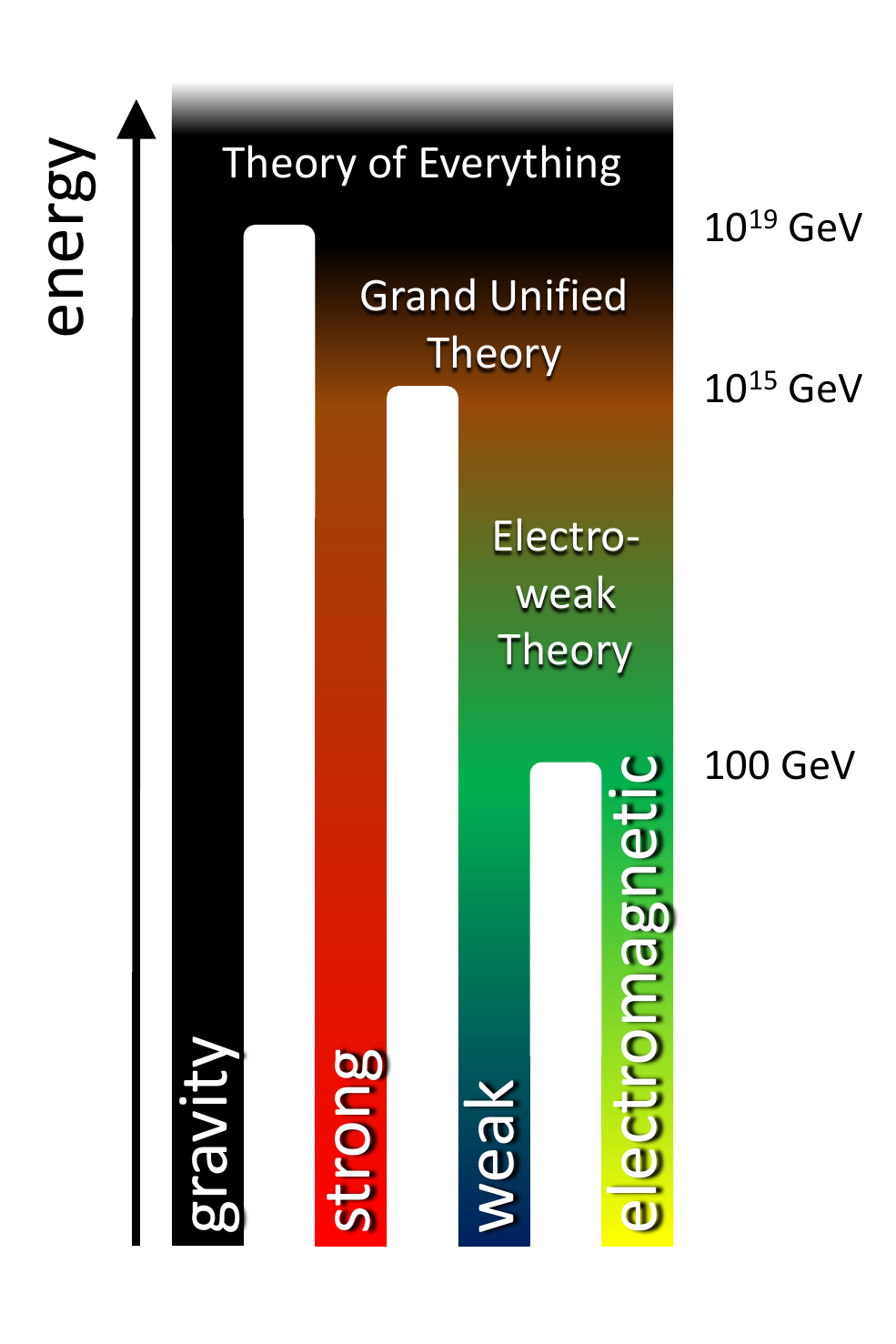}
\end{center}
\caption{\label{fig:GUT}Unification of fundamental forces. There is strong experimental evidence that at energies above $100~{\rm GeV}$ the electromagnetic and weak nuclear forces are described by a unified electroweak theory. It is possible that this is unified with strong nuclear forces to a Grand Unified Theory (GUT) above $10^{15}~{\rm GeV}$. Finally, it is known that at energies above the {\it Planck scale}, $10^{19}~{\rm GeV}$, the gravitational force becomes so strong that it needs to be included in the theory, and it is believed that it will be unified with all the elementary particle forces into a Theory of Everything, which may be a superstring theory or the M-theory~\cite{Duff:1996aw}.
}
\end{figure}

In particular, progress was being made to unify electromagnetism and the weak nuclear force into one {\it electroweak} force (see Fig.~\ref{fig:GUT}). The theory that achieved this was developed by Sheldon Glashow, Steven Weinberg, Abdus Salam and others, and has two different gauge symmetries known as U(1) and SU(2). 
The U(1) gauge symmetry has the same structure as the gauge symmetry of Maxwell's equations, but the SU(2) symmetry is somewhat more complicated. In particular, in U(1) the order in which one makes two gauge transformations does not but in SU(2) it does. Because of this, the former is known as an Abelian symmetry, and the latter as a non-Abelian symmetry.
At low energies, these symmetries are ``broken'' because of a hypothetical {\it Higgs field}, which has a non-zero value in the vacuum and behaves essentially like a quantum wave function spread over the whole space. Just like the complex phase of the wave function, the Higgs field is sensitive to the values of the vector potentials corresponding to the two gauge symmetries. Although the theory itself still has the gauge symmetry, the vacuum state with the non-zero Higgs field is no longer symmetric, and therefore only certain gauge transformations are allowed. These turn out to correspond exactly to the U(1) gauge symmetry of Maxwell's equations, and therefore they gives rise to the electromagnetic forces. As a by-product the theory predicted the existence of some new particles, $W$ and $Z$ bosons which were discovered in 1980s and the Higgs boson which the Large Hadron Collider (LHC) at CERN is currently trying to find.

In 1974 Gerard 't~Hooft~\cite{'tHooft:1974qc} and Alexander Polyakov~\cite{Polyakov:1974ek} 
considered a slightly different variant of this theory, which had been proposed earlier by Georgi and Glashow~\cite{Georgi:1972cj}. 
Remarkably, 't~Hooft and Polyakov found that this theory would necessarily contain both electric and magnetic charges.

\begin{figure}
\begin{center}
 \includegraphics[width=5cm]{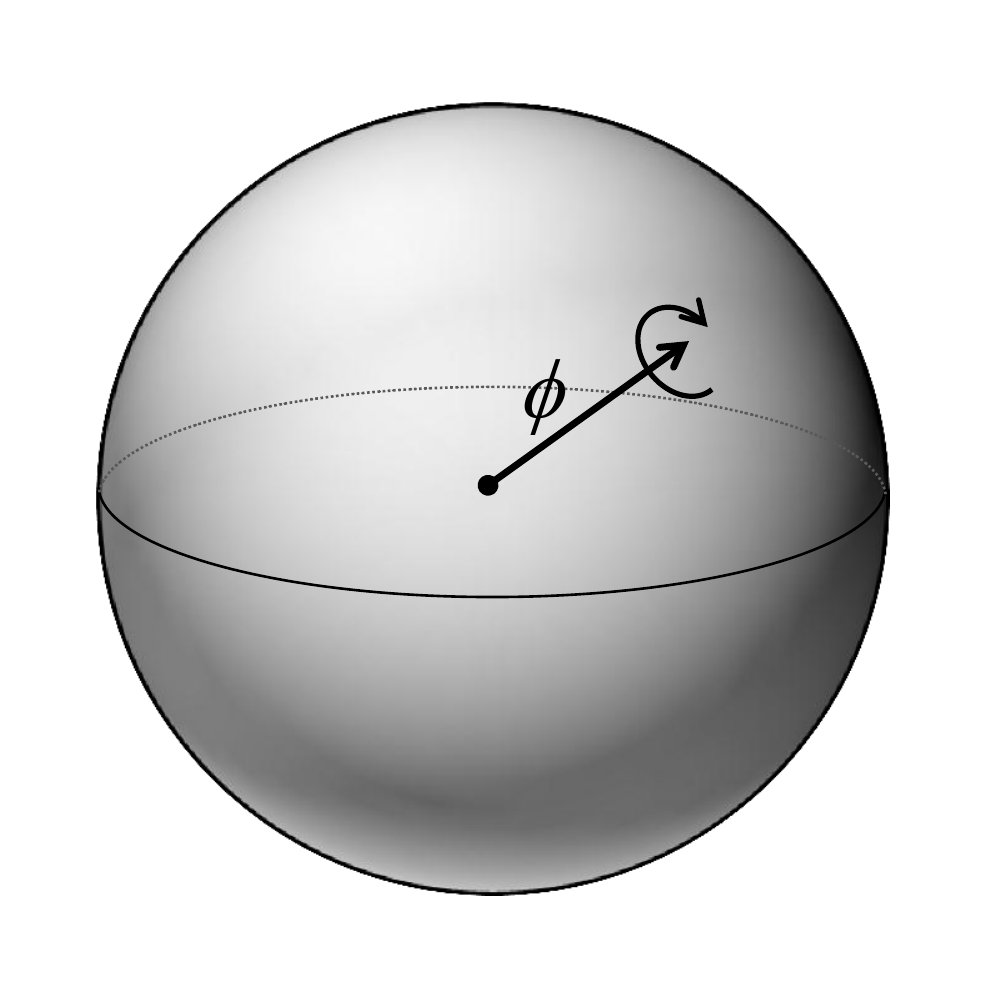}
\end{center}
\caption{\label{fig:SSB}Spontaneous symmetry breaking. The Georgi-Glashow model of electroweak unification has an SO(3) gauge symmetry which correspond to rotations in the three-dimensional internal space. However, the Higgs field $\phi$ has a non-zero length, and therefore its possible values lie on a sphere in the internal space, corresponding to different possible vacuum states. Because of the SO(3) symmetry, all vacuum states are identical, but once the vacuum state has been chosen, only rotations around the Higgs field axis are possible. Therefore the full SO(3) symmetry is spontaneously broken to a smaller U(1) symmetry, which gives rise to electromagnetism.} 
\end{figure}

The Georgi-Glashow model is somewhat simpler than the Glashow-Weinberg-Salam SU(2)$\times$U(1) electroweak theory. It has only one gauge symmetry, a non-Abelian symmetry known as SO(3).
The Higgs field consists of three components, which are real numbers. We can therefore think of it as a three-component vector, although it is a vector in the abstract ``internal'' space rather real space. The gauge transformations correspond to rotating this internal space, and the gauge symmetry means that such a rotation does not change the laws of physics. 

In the vacuum state, the Higgs field has a non-zero value: The length of the vector is fixed, but it can point in any direction, so there is a sphere of different vacuum states (see Fig.~\ref{fig:SSB}). Once the direction is fixed, only rotations around that axis are allowed, so the original SO(3) symmetry is said to be spontaneously broken. The remaining rotations correspond to the U(1) gauge symmetry of Maxwell's equations, and therefore they give rise to electromagnetism.
Furthermore, the requirement that a charged particle would have to behave consistently under any rotation of the whole three-dimensional internal space, not just those around the Higgs field vector axis, gives rise to a condition that the electric charge has to be quantised in units of some elementary charge $q$.

\begin{figure}
\begin{center}
 \includegraphics[width=7cm]{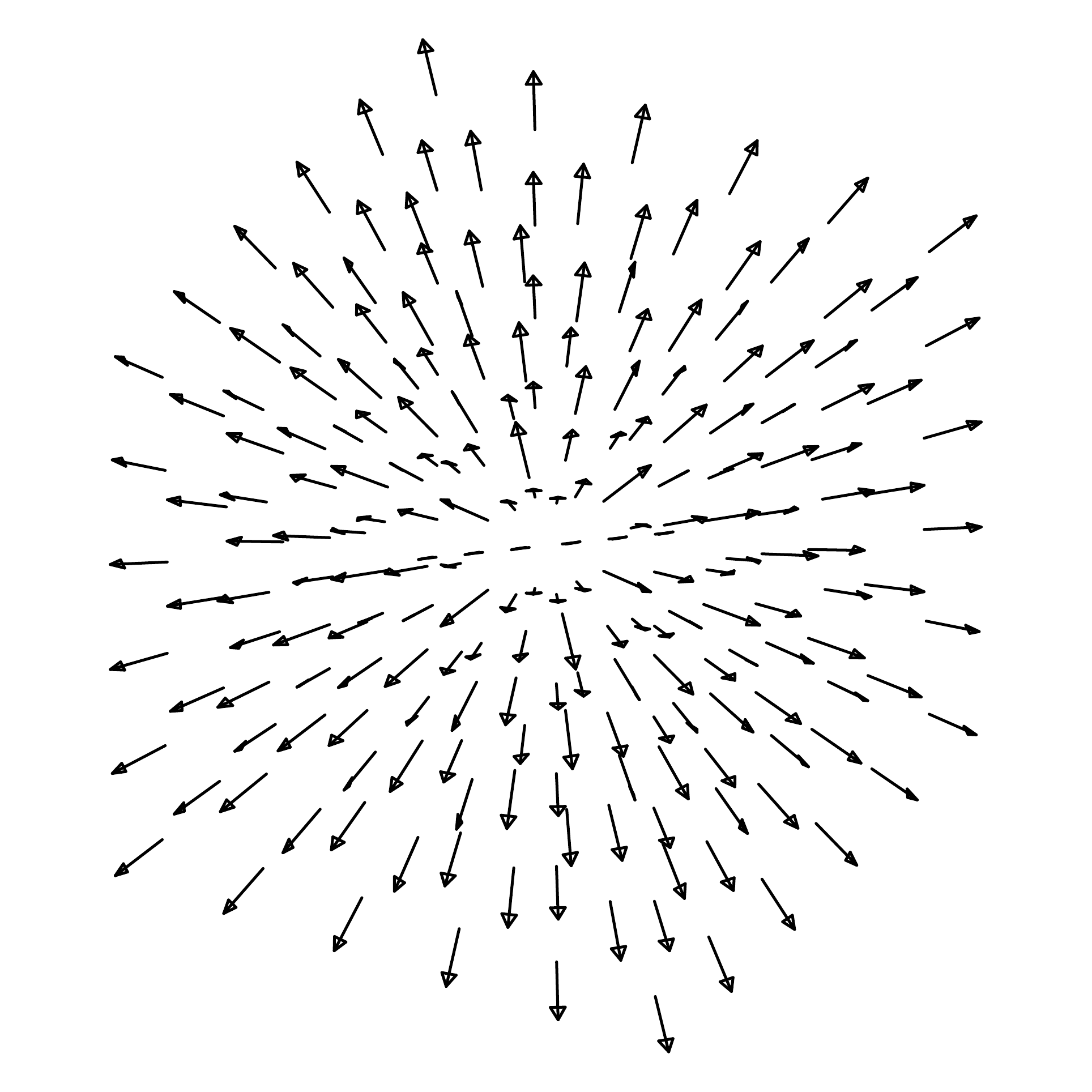}
\end{center}
\caption{\label{fig:hedgehog}The hedgehog configuration. In the 't~Hooft-Polyakov monopole solution, the Higgs field has a fixed length, but its direction (indicated by the arrows) is different in different directions, in such a way that it always points away from the origin. This configuration cannot be turned continuously into the uniform vacuum state, so it is topologically stable.
In order for the field to be continuous, it cannot be in the vacuum state at the origin, and therefore there is a localised lump of energy (in other words, a particle) at the origin.}
\end{figure}

't~Hooft and Polyakov considered what happens if the direction of the Higgs field is not the same everywhere. In most cases, the field would quickly find its way to the vacuum state. However, they found that this would not happen to the so-called hedgehog configuration, which one can visualise by imagine that the directions in the internal space correspond to directions in real space. The hedgehog configuration would then be the case in which the Higgs field vector points away from the origin everywhere (see Fig.~\ref{fig:hedgehog}). This field configuration is an example of a topological defect or a topological soliton~\cite{solitons}: It is stable because it is impossible to turn it continuously into the uniform vacuum state.

Far from the origin, the Higgs field is almost uniform, so it is almost in the vacuum state. However, near the origin its direction varies rapidly and at the origin its length has to go to zero, which means that it will have non-zero energy. Therefore, the hedgehog configuration appears physically as a lump of energy localised in a small volume near the origin. 
Because according to the theory of relativity, mass is energy, this means that it is essentially a massive particle.
Furthermore, because the Higgs field vanishes at the origin, the full SO(3) gauge symmetry is unbroken inside the particle.

However, the most striking result is that the particle has a magnetic charge. Because electromagnetism is generated by rotations around the Higgs field vector, 
one can calculate the magnetic field in the hedgehog configuration~\cite{'tHooft:1974qc}, and it turns out to be just the field of a magnetic monopole with magnetic charge
\begin{equation}
g=\frac{4\pi}{q},
\end{equation}
which is twice the minimum allowed by the Dirac quantisation condition (\ref{equ:Diracquant}). 

The mass of these 't~Hooft-Polyakov monopoles would be roughly $100~{\rm GeV}$, determined by the energy scale associated with weak nuclear forces.
However, experimental results quickly ruled out the Georgi-Glashow model as a theory of electroweak unification, and the successful SU(2)$\times$U(1) theory does not allow similar magnetic monopole solutions.

However, Georgi and Glashow noticed very soon that their theory could be modified to do something even more ambitious, namely to unify the SU(2)$\times$U(1) electroweak theory with strong nuclear forces into one Grand Unified Theory (GUT)~\cite{Georgi:1974sy}, which would describe all known elementary particle forces except gravity (see Fig.~\ref{fig:GUT}). They achieved this very elegantly using only one non-Abelian gauge symmetry known as SU(5), and two Higgs fields. At very high energies of around $10^{15}~{\rm GeV}$, one of the Higgs fields would break the GUT symmetry into three pieces, the SU(2) and U(1) symmetries of the electroweak theory and a further SU(3) symmetry that describes strong nuclear forces, and the second Higgs would then play exactly the same role as in the electroweak theory. 

Because the structure of the SU(5) GUT was so similar to the Georgi-Glashow electroweak theory, it also predicted 't~Hooft-Polyakov monopoles, but their mass would now be much higher, around $10^{16}~{\rm GeV}$ (or $10~{\rm ng}$ in SI units). They also have the interesting property that they can catalyse nucleon decay~\cite{Callan:1982ac,Rubakov:1982fp,Rubakov:1988aq}. According to the Standard Model of particle physics, the proton is a stable particle because and cannot decay. In the grand unified theory this is no longer true, but the proton still has a very long lifetime because the decay requires a quantum tunnelling process through a very high potential barrier. However, as we saw earlier, the full SU(5) symmetry is unbroken inside the monopole and therefore the potential barrier disappears. This means that when a proton or a neutron gets in contact with a GUT monopole, they can decay very fast. 

In fact, the existence of magnetic monopoles is not specific to the SU(5) GUT. Instead, it is an unavoidable consequence of grand unification: Any theory that would describe all strong and electroweak forces by one grand unified interaction would have 't~Hooft-Polyakov monopoles~\cite{Preskill:1984gd}. The reason for this lies in a branch of mathematics known as topology, which studies whether one function can be continuously deformed to another.

The way to demonstrate that the 't~Hooft-Polyakov hedgehog configuration is stable is to consider a sphere around the origin. In the Georgi-Glashow model, the Higgs field is a three-component vector, and because its length is fixed, the set of possible values it can take also a form a sphere in the internal space. In the hedgehog solution, the field values on a sphere in real space wrap once around the sphere in the internal space, and this cannot be continuously deformed into the vacuum state, in which the field values do not wrap around the sphere, and this makes the configuration stable.
In topology, homotopy theory tells precisely when stable configurations like this exist, and one can show that whenever a gauge symmetry is broken by a Higgs field in such a way that a U(1) gauge symmetry remains, stable hedgehog configurations exist~\cite{Kibble:1976sj,Preskill:1984gd}. Because a U(1) gauge symmetry is needed for electromagnetism, this implies that any GUT that unifies electromagnetism with other forces into one gauge symmetry would always have stable hedgehogs, i.e., 't~Hooft-Polyakov monopoles.

This was a remarkable finding, because there were (and still are) many theoretical reasons to expect grand unification. First of all, the Standard Model of particle physics has a fairly complex structure consisting three different gauge symmetries and a wide range of different matter particles each with their own properties, which gets a simple explanation in grand unified theories. The strengths of the three interactions are also such that that they coincide when extrapolated to the energy of $10^{15}~{\rm GeV}$. However, even if grand unification is valid as a general concept, we cannot be sure that the Georgi-Glashow SU(5) GUT is the correct theory. It is therefore important that the prediction of magnetic monopoles is general, and not specific to this particular theory.

It is obviously almost impossible to test grand unified theories with accelerator experiments, because that would require energies around $10^{15}~{\rm GeV}$, which is one trillion times higher than what the LHC can reach. It is therefore exciting that magnetic monopoles can give a potential way to study them experimentally. However, as we saw before, grand unified theories also predict that the proton can decay. This means that hydrogen atoms, which have a single proton as their nucleus should have finite lifetime. Because every water molecule contains two hydrogen atoms, one should be able to detect this by observing a large enough tank of water. At some point a proton should decay in the tank, creating a tiny flash of light, which can be detected. However, no proton decays have ever been seen in these experiments. This rules out the original Georgi-Glashow SU(5) GUT, but there are other GUT which predict a much longer proton lifetime and and which are therefore still compatible with experiments.

These days, the focus in high energy physics has shifted from grand unified theories to even more ambitious ``theories of everything'' such as superstring theory, which aim to include also gravity. The structure of these theories is different because all four forces are unified at the same energy and because they are not based on fields but on other types of objects such as strings or membranes. Therefore the argument for the existence of 't~Hooft-Polyakov monopoles does not apply in the same form. However, they still generally predict the existence of magnetic monopoles~\cite{Duff:1994an}.

\section{Monopoles in Cosmology}
\label{sec:cosmo}
If the fundamental laws of physics allow the existence of magnetic monopole particles, then it is likely that they would have been produced in the Big Bang or shortly afterwards in the early universe~\cite{Zeldovich:1978wj,Preskill:1979zi}. Therefore, with a good enough understanding of the production mechanism and the early universe, one should be able to calculate how many monopoles there should be in the universe today. However, even very simple estimates show that the predicted is far too high to be compatible with observations~\cite{Preskill:1979zi}. This is known as the {\it monopole problem}. 

To see this, let us start by looking at how the monopoles may have been produced in the early universe. The basic idea is known as the {\it Kibble mechanism} after Tom Kibble who discovered it in 1976~\cite{Kibble:1976sj}, and although one can argue that it is not strictly speaking applicable to grand unified theories~\cite{Rajantie:2002dw}, the main conclusions are still valid~\cite{Rajantie:2001ps,Rajantie:2003wi}.

\begin{figure}
 \begin{center}
  \includegraphics[width=8cm]{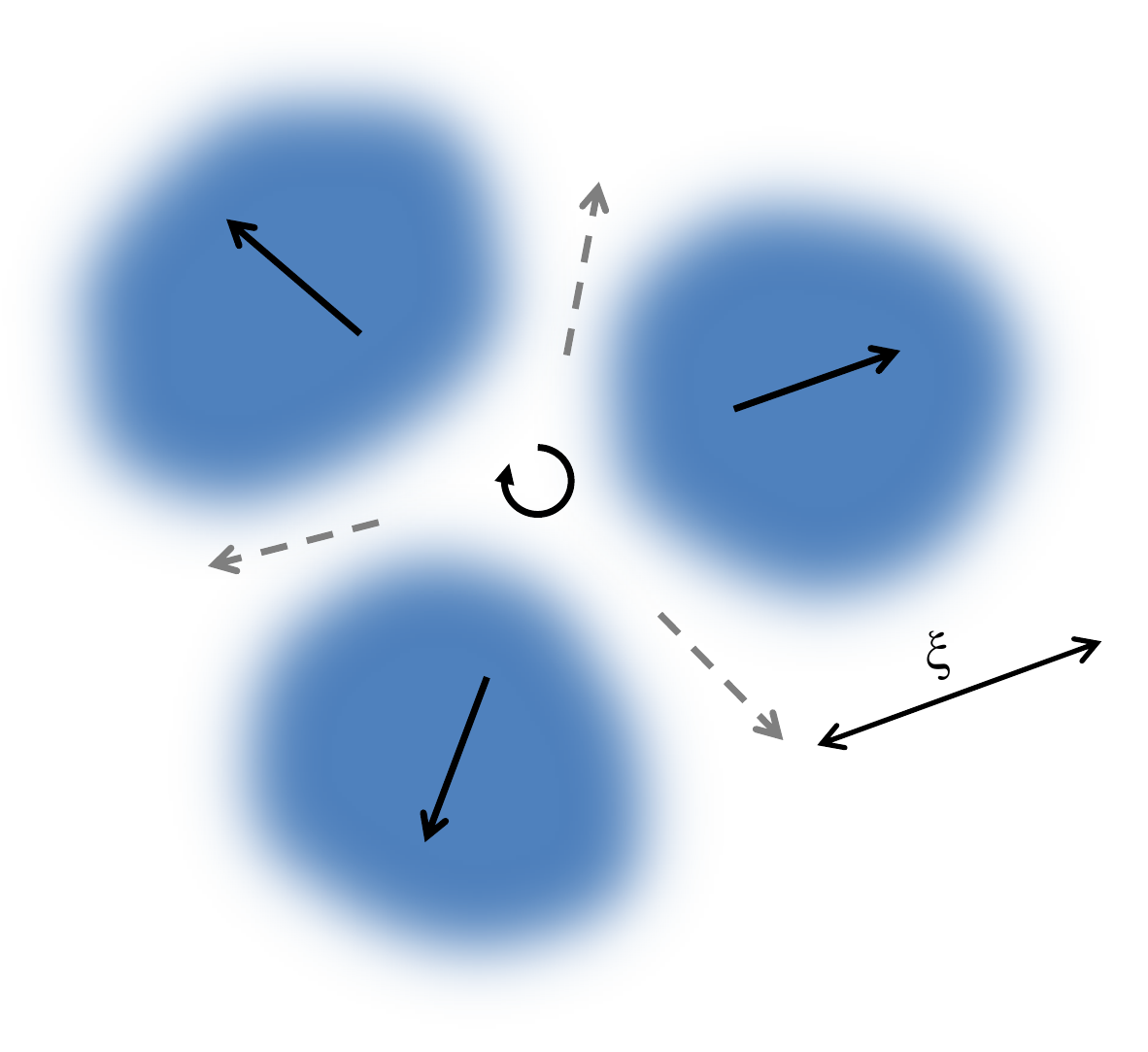}
 \end{center}
\caption{\label{fig:kibble}A two-dimensional example of the Kibble mechanism. The Higgs field is random at distances longer than the correlation length $\xi$. Therefore we can imagine that the universe consists of domains of size $\xi$, each with a random but more or less uniform field direction (solid arrows). The field interpolates continuously between any pair of two domains (dashed arrows), but where three domains meet, it may not be able to do that without vanishing in the middle. In that case a topological defect (a vortex) is formed. Similarly, in three dimensions, monopoles can be formed where four domains meet.
}

\end{figure}

For simplicity, let us consider the SO(3) Georgi-Glashow model in which the Higgs field can be though of as a three-component vector.
The early universe was very hot, and if it was hot enough the GUT symmetry would have been initially unbroken, and the Higgs field was zero. When the universe expanded and cooled down, it went through a phase transition to the phase with the broken gauge symmetry when the universe was very young, only $10^{-35}~{\rm s}$ old. In this phase transition, the Higgs field became non-zero, and because it is a vector it has to have a direction. In order to minimise the energy, the direction would have to be the same everywhere.

However, the transition took place in a finite time. Remember that according to relativity, information cannot travel faster than the speed of light. If you consider two points in space that are so far apart that information from one cannot reach other before the transition is completed, then those two points have to choose the direction of the Higgs field independently of each other. 
Because of the symmetry, every direction is equally likely. Therefore it follows that after the transition the Higgs field will be uniform on short scales, up to the distance $\xi$ that light was able to travel during the transition, but random and uncorrelated at longer distances. 
Roughly speaking, we can imagine that the universe consists of domains of size $\xi$, in each of which the direction of the Higgs field is uniform but random.
In the early universe, this correlation distance $\xi$ cannot have been longer than the size of the particle horizon, which is roughly given by the age of the universe or the inverse of the expansion rate (or Hubble rate) $H$ at the time.

On the other hand, the Higgs field has to be continuous, so where two domains meet, it will interpolate smoothly between them. However, consider a point between four domains. It is possible that, when the field interpolates between each pair of them, it ends up wrapping around the sphere in the internal space. In that case, the field cannot continuously interpolate between all four domains without vanishing in the middle, and a monopole or an anti-monopole (a monopole with negative magnetic charge) is formed (see Fig.~\ref{fig:kibble}). The probability that the directions in the four domains are such that this happens is independent of their size or any other parameter, so it is simply some constant number which can be calculated in principle but which is clearly going to be not much less than one. Therefore one concludes that roughly one monopole or anti-monopole is is produced per domain, or per particle horizon~\cite{Kibble:1976sj}. Because the particle horizon was very short in the very early universe, when the GUT phase transition would have taken place, this leads to a very high density of monopoles and anti-monopoles.

Nevertheless, the magnetic monopoles carry initially only a very small fraction of the total energy, most of which is in the form of radiation. 
Because both monopoles and anti-monopoles are produced in equal numbers, they can initially meet and destroy each other reducing their number density. However, after a while those processes stop, and after that the number density of monopoles decreases only because of the expansion of the universe. John Preskill calculated in 1979 that today the number density of magnetic monopoles would be comparable to the number density of protons and neutrons~\cite{Preskill:1979zi,Preskill:1984gd}. The problem is that the mass of a monopole, at roughly $10^{16}~{\rm GeV}$, is fifteen orders of magnitude higher than the mass of a proton or neutron. It is therefore absolutely clear that the prediction cannot be valid.

One possible solution to this monopole problem is that magnetic monopoles simply do not exist. However, in 1980 Alan Guth proposed an alternative explanation in the form of a theory called inflation~\cite{Guth:1980zm}. According to this theory, the universe expanded at an accelerating rate in its very early stages. If this happened after the GUT phase transition (or during it as Guth originally thought), it could potentially dilute the monopole density to an acceptable level. With a sufficiently long period of acceleration, inflation would also solve
two other major puzzles in Big Bang cosmology, known as the horizon and flatness problems. Furthermore, a more detailed calculation shows that inflation also
explains the origin of primordial density perturbations which acted as the initial seeds for the formation of structures such as galaxies and galaxy clusters.
There same perturbation can be observed more directly as temperature fluctuations in the cosmic microwave background radiation. Measurements of these temperature fluctuations, by the WMAP satellite~\cite{Komatsu:2010fb} and other experiments, have shown that they agree astonishingly well with the predictions of the inflationary theory.

In spite of its great successes, inflation is more a general idea than a specific theory. There are a large number of theoretical models that give rise to enough inflation to solve the monopole, horizon and flatness problems and can produce perturbations that are compatible with observations.
In most of these models, inflation takes place at energies well below the GUT scale. Therefore, if monopoles were formed at all, that would have happened before inflation. Inflation would then have solved the monopole problem by diluting their number density down to a small enough level. 

However, there are some models in which GUT or intermediate mass monopoles are formed in a phase transition at the end of inflation or shortly afterwards~\cite{Rajantie:2003wi}. In those cases it will be important to understand the precise mechanism by which the monopoles form and the way they evolve afterwards.

\section{Monopole Searches}
Following the theoretical prediction by 't~Hooft and Polyakov, there have been many attempts to detect magnetic monopoles in experiments. These fall in three categories: One can try to produce them in accelerator experiments such as the LHC, one can try to find existing magnetic monopoles either in cosmic rays or trapped in materials, or one can look for indirect signs of magnetic monopoles in astronomical observations. 

For direct searches there are three primary methods: The first is to use a superconducting ring. If a magnetic monopole passes through the ring, the changing magnetic field induces a current, which can be measured. The change in the current also gives the strength of the magnetic charge very accurately. The second method relies on the fact that magnetic monopoles are very highly ionising: Because of their strong magnetic charge, a magnetic monopole travelling through matter would easily strip electrons off atoms, leaving a clear track in detector plates. Because of the form of the Lorentz force (\ref{equ:LorentzE}), this effect is velocity dependent, which allows one to distinguish a monopole from an strong electric charge. The third  approach is to look for signs of nucleon decay catalysed by monopoles travelling through matter~\cite{Rubakov:1982fp,Errede:1983mb}.

The following is a brief summary of different experimental searches. More details can be found in Refs.~\cite{Nakamura:2010zzi,Giacomelli:2011re}.

\subsection{Accelerator Searches}
In principle, magnetic monopoles should be produced in particle accelerator experiments if the collision energy is sufficiently high, higher than $2Mc^2$. For GUT monopoles the required energy is at least one trillion times higher than the energies available at the Large Hadron Collider (LHC). Therefore, it is unrealistic to expect that they could be produced in any foreseeable particle accelerators. 

However, we really only know the laws of physics up to the electroweak scale, which is at around $100~{\rm GeV}$, and therefore it is perfectly possible that there are magnetic monopoles that are much lighter than GUTs would predict. These are called intermediate-mass monopoles, and if they are light enough, they could very well be produced at the LHC, although there are huge theoretical uncertainties in their predicted production rate~\cite{Drukier:1981fq} (see Section~\ref{sec:QFT} for more discussion).

Looking for magnetic monopoles is very different from looking for most other hypothetical particles. Exotic particles are usually very difficult to detect because they have a very short lifetime. The particle decays very quickly into known particles, and therefore one has to analyse the particles produced in the collision very carefully to see if some of them might have been created by the hypothetical new particle. This is very difficult, because it is much more likely that the same known particles were produced in other processes.

In contrast, magnetic monopoles are stable. A monopole can only be destroyed when it meets an anti-monopole. Therefore, once created in an experiment, the monopole will not decay. Because of its magnetic charge, the monopole will also interact relatively strongly with the electromagnetic field. This interaction is well understood, and also very different from any known particles, and therefore there would be little risk of confusion.

Because of these differences, the main LHC experiments are not well suited for looking for magnetic monopoles. Instead,
a dedicated experiment called MoEDAL~\cite{Pinfold:2010zza} is being constructed for that purpose. Compared with the other LHC experiments, MoEDAL is very simple: It consists of plastic nuclear track detector sheets placed around the LHCb experiment. If monopoles are produced in collisions, they will fly through the sheets, leaving tell-tale marks on them which can be detected when the sheets are removed and analysed. Monopoles have also been searched for in different ways in other accelerators in the past, such as Tevatron, LEP and HERA, but with no success. This excludes monopoles with mass less than roughly $1~\rm {TeV}$~\cite{Fairbairn:2006gg}.

In principle, quantum field theory also allows the possibility that a ``virtual'' pair or two oppositely charged monopoles are produced and destroy each other after a very short time. This process is possible even when the collision energy is not quite high enough to produce two real monopoles. It is much harder to look for virtual monopoles, but there have been some attempts~\cite{Acciarri:1994gb,Abbott:1998mw}

\subsection{Direct searches}
Instead of trying to produce monopoles in an experiment, one can also try to look for monopoles that already exist in the universe. Because monopoles are stable, even monopoles created in the very early universe should still be around. Obviously, how easy it is to find them depends on how many of them there are, or more precisely on their flux $F$. By this we mean the number of monopoles that would hit a unit area per unit time per solid angle. The flux is conventionally measured in units of ${\rm cm}^{-2}{\rm s}^{-1}{\rm sr}^{-1}$, where ${\rm sr}$ stand for {\it steradian} and is the unit of solid angle. The solid angle of a whole sphere is $4\pi{\rm sr}$. Therefore the number of monopoles hitting a sphere of radius $R$ in time $t$ is equal to $(4\pi{\rm sr})\times FRt$. This clearly depends not only on the number of monopoles but also on their velocity.

\begin{figure}
 \begin{center}
(a)\hspace*{10cm} \\  \includegraphics[width=8cm]{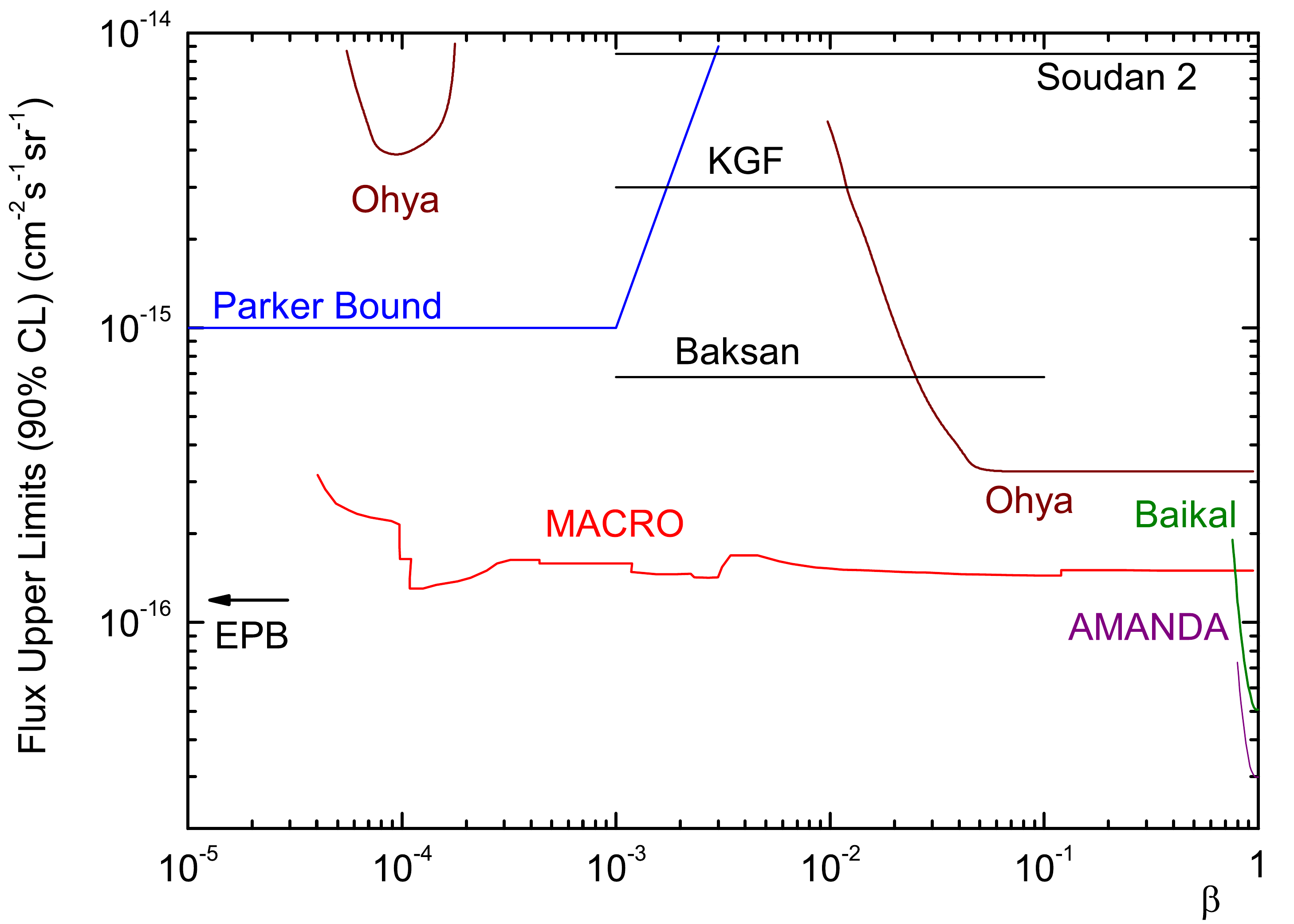}\\
(b)\hspace*{10cm} \\  \includegraphics[width=8.5cm]{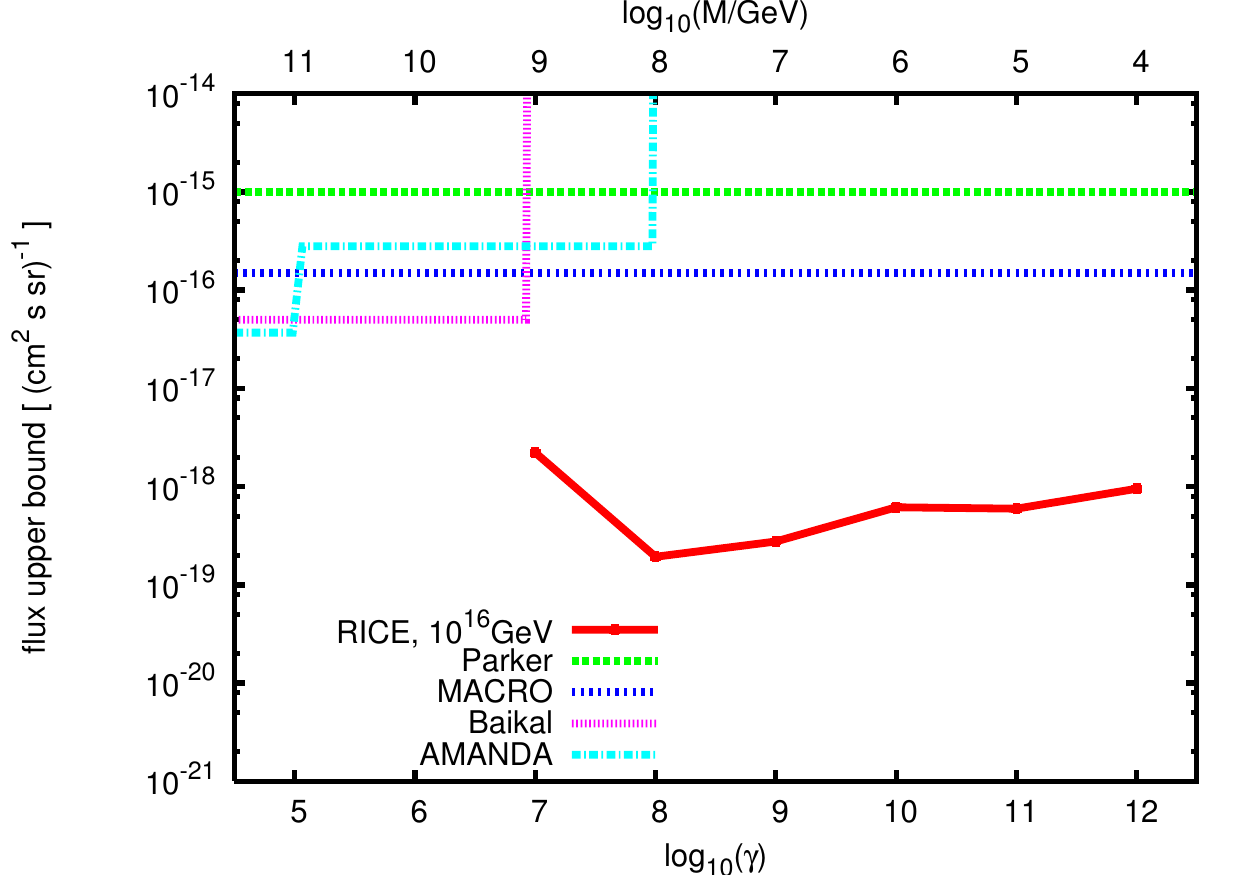}
 \end{center}
\caption{\label{fig:RICE}Upper bounds for the flux of 
(a) GUT monopoles (From Ref.~\cite{Giacomelli:2011re}), and
(b) intermediate mass magnetic monopoles (From Ref.~\cite{Hogan:2008sx}),
from several experiments as well as the astrophysical Parker bound. In (a) the x axis $\beta$ is the velocity of the monopoles in units of the speed of light, $\beta=v/c$.
}
\end{figure}

There have been many attempts to look for magnetic monopoles in cosmic rays, which
are particles hitting the Earth from space. The higher the flux, the more often a monopole cosmic ray would hit a detector.
Some early experiments appeared to show promising evidence for them. In 1973, a cosmic ray balloon experiment showed a particle track that appeared to be consistent with a magnetic monopole~\cite{Price:1975zt}, but it was soon accepted that it was more likely to have been a platinum nucleus~\cite{Alvarez:1975gm}. On Valentine's day in 1982, a laboratory experiment using a superconducting ring showed a sudden jump in the current by exactly the amount that a magnetic monopole passing through the ring would generate~\cite{Cabrera:1982gz}, and another event was seen in a similar experiment at Imperial College in 1985~\cite{Caplin:1986kw}. However, because later experiments have not been able to reproduce them, they are believe to have been caused by other effects.
Nevertheless, they motivated physicists to design larger experiments which were able to put stronger limits on the flux of the magnetic monopoles.
Furthermore, these experiments were also able to look for other types of particles in cosmic rays and they paved way for current dark matter searches.

The strongest bounds for GUT monopoles come from an underground experiment called MACRO (Monopole, Astrophysics and Cosmic Ray Observatory) which was based in Gran Sasso, Italy and operated from 1989 to 2000. It was specially designed to detect magnetic monopoles, and contained different types of detectors using both ionisation and nucleon decay methods, with a total effective surface area of roughly $10000~{\rm m}^2$. It did not detect any monopoles, and from this one can conclude that the monopole flux cannot be higher than roughly $F\lesssim 10^{-16}{\rm cm}^{-2}{\rm s}^{-1}{\rm sr}^{-1}$~\cite{Ambrosio:2002qq} (see Fig.~\ref{fig:RICE}(a)). Several other experiments such as  AMANDA~\cite{Abbasi:2010zz} and Baikal~\cite{Antipin:2007zz} have produced similar bounds later. For lighter, intermediate-mass monopoles the RICE experiment (Radio Ice Cherenkov Experiment) at the South Pole has given an even stronger bound of roughly $F\lesssim 10^{-18}{\rm cm}^{-2}{\rm s}^{-1}{\rm sr}^{-1}$~\cite{Hogan:2008sx} (see Fig.~\ref{fig:RICE}(b)). 

Because the Earth has been bombarded by cosmic rays for billions of years, monopoles could have well been trapped in matter on Earth. 
There have been attempts to find magnetic monopoles trapped in materials like moon rock, meteorites and sea water, but with no success~\cite{Jeon:1995rf}.

\subsection{Astrophysical bounds}
If magnetic monopoles exist, they would produce astrophysical effects, 
which we can hope to detect in astronomical observations.
Perhaps the most significant effect is that magnetic fields would accelerate them, generating magnetic currents which would drain energy from the field.
The rate at which the energy of the magnetic field would decrease depends on the monopole flux. If the flux was very high, we would not expect to find any large-scale magnetic fields in space because they would have dissipated away. However, observations show that there is a magnetic field of roughly $3\mu G$ in our galaxy, from which we can conclude that the monopole flux cannot be very high. This is known as the {\it Parker bound}~\cite{Parker:1970xv,Turner:1982ag}, 
The strength of this effect depends also on the mass of the monopoles. If they are very heavy, around $10^{17}~{\rm GeV}$, their motion is dominated by gravitational forces rather than magnetic fields, and therefore the bound is weak (see Fig.~\ref{fig:RICE}). On the other hand, for lighter monopoles one finds a stronger bound 
\begin{equation}
F\lesssim 10^{-15}{\rm cm}^{-2}{\rm s}^{-1}{\rm sr}^{-1}.
\end{equation}

An even simpler bound follows from the total mass of the magnetic monopoles in the universe. From astronomical observations~\cite{Komatsu:2010fb}, we know that matter particles make up approximately $23\%$ of the total energy in the universe, and roughly one third of this can be accounted for by known particles. In principle, some or all of the remaining dark matter could be magnetic monopoles. If they are light enough, less than $10^{17}~{\rm GeV}$, they would not be bound to galaxies by gravity, and therefore their density would be more or less uniform. From the observed dark matter density, one obtains an upper bound
\begin{equation}
F\lesssim 10^{-16}(10^{16}~{\rm GeV}/M)(v/10^{-3}c){\rm cm}^{-2}{\rm s}^{-1}{\rm sr}^{-1},
\end{equation}
 where $v$ is the average speed of the monopoles and $M$ is their mass. Because of the mass dependence, this bound is only relevant for heavy monopoles, such as GUT ones, but for them it is comparable to other bounds.

\section{Monopoles and Quantum Field Theory}
\label{sec:QFT}
Even though magnetic monopoles have never been discovered, and therefore we do not know if they really exist, they have still been 
extremely useful for theoretical physics. We already saw in Section~\ref{sec:cosmo} how they were the motivation for inflation, which has later been confirmed by other observations. They have also played an equally important role in understanding quantum field theories.

Attempts to include magnetic monopoles have forced physicists to look at quantum field theories from a new perspective and to develop new calculation techniques. Among the early examples is Schwinger's formulation of quantum electrodynamics with magnetic charges in 1966~\cite{Schwinger:1966nj}, but this direction became much more important with the discovery of 't~Hooft-Polyakov monopoles in 1974. The conventional interpretation of quantum field theories was that particles correspond to field quanta, or quantised waves, so that each particle species would correspond to one particular field. For example, photons are the quanta of the electromagnetic field. However, 't~Hooft-Polyakov monopoles are localised field configurations, which appear in the theory in addition to all the field quanta. These days analogous topological solitons~\cite{solitons} play a central role especially in superstring theory.

Describing solitons such as 't~Hooft-Polyakov monopoles in quantum field theory gives rise to new challenges. The conventional way to do calculations in quantum field theory, known as perturbation theory, is designed for field quanta, and it cannot be used to topological solitons. Furthermore, it assumes that the quanta interact with each other weakly. For example, the strength of the interactions between electrons and photons is given by the electron charge $e$, or perhaps more precisely by the fine structure constant $\alpha=e^2/4\pi\approx 1/137$. This is a small number, and therefore the interaction is weak, and perturbation theory works. The problem is that the Dirac quantisation condition (\ref{equ:Diracquant}) tells that the magnetic charge of the magnetic monopole is very large. The analogous magnetic fine structure constant would be
$\alpha_{\rm M}=g^2/4\pi=1/\alpha\approx 137$, which is not a small number. Therefore the magnetic interaction between two monopoles is actually very strong, and perturbation theory is generally not applicable.

\begin{figure}
 \begin{center}
  \includegraphics[width=7cm]{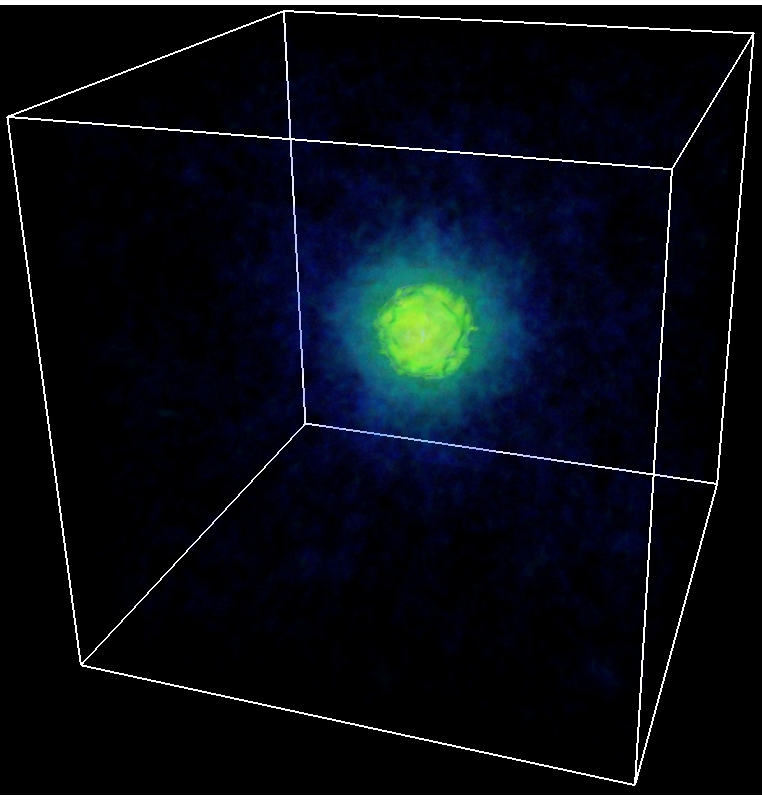}
 \end{center}
\caption{\label{fig:lattice}A 't~Hooft-Polyakov monopole configuration in a lattice field theory simulation. This figure was produced using VAPOR~\cite{Clyne:VDA:2005,Clyne:NJP:2007}.
}
\end{figure}

The main alternative to perturbation theory is to do numerical lattice field theory simulations. These simulations are widely used to study quantum chromodynamics (QCD), the theory of the strong nuclear force. As the name suggests, this interaction is not weak either. However, because magnetic monopoles and other solitons are very different in nature from the particles involved in QCD, the techniques used in lattice QCD are not always directly applicable. Instead, one needs develop new techniques to calculate the properties of magnetic monopoles, such as their mass and interactions with other particles.
This is particularly important because there are experimentally important quantities which one cannot calculate reliably using other techniques, for example the effects of virtual monopole-antimonopole pairs. Most importantly, semiclassical calculations suggest that the pair production rate of magnetic monopoles in particle collisions is suppressed by a massive factor which would make it practically impossible to find them in accelerator experiments such as the LHC~\cite{Drukier:1981fq}, and it is crucial to check whether this is also true in quantum theory. There has been significant progress recently in developing these methods, and they have been used to calculate the mass and the interaction form factors of magnetic monopoles~\cite{Davis:2000kv,Rajantie:2005hi,Rajantie:2011nq}. Fig.~\ref{fig:lattice} shows a visualisation of a 't~Hooft-Polyakov monopole in such a lattice field theory simulation.

Magnetic monopoles may also be able to help us understand the physics of QCD better.
The theory is defined in terms of two types of particles: quarks which are matter particles, and gluons which mediate the strong force. However, neither quarks nor gluons have been seen as free particles in experiments. Instead, we know that they are confined into protons, neutrons and other hadron particles, which are all made up of several quarks. If we tried to pull the quarks apart, we would feel a constant force keeping them together as if they were connected by a string.
This phenomenon is called {\it confinement}, and 
we know from lattice field theory simulations that QCD can describe it correctly, but there is no rigorous proof becuase we do not have any reliable ways of doing analytic calculations with the theory.

In 1976, Gerard 't~Hooft~\cite{tHooft1976} and Stanley Mandelstam~\cite{Mandelstam:1974pi} suggested that confinement could be explained by magnetic monopoles.
Note first that under certain circumstances (typically low temperature) some materials become superconducting. This means that electric currents can flow without any resistance, but also that magnetic fields are repelled. If one could put a magnetic monopole inside a superconductor, its magnetic field would be confined into a string known as an Abrikosov flux tube. A monopole and an anti-monopole would therefore we connected by a string.

't~Hooft and Mandelstam realised that confinement looks looks very much like a dual version of this in the sense of the electric-magnetic duality (\ref{equ:duality}), i.e., when we swap electric and magnetic fields and charges.
Because superconductivity is caused by the behaviour of electrically charged particles, confinement could therefore be explained by magnetic monopoles behaving in a similar way. These monopoles would not be the ``real'' GUT monopoles, but analogous mathematical solutions in QCD.
It is difficult to make this idea precise in QCD, but in some simpler theories,
people have been able to show rigorously that it is correct and that confinement is indeed caused by magnetic monopoles~\cite{Polyakov:1976fu,Seiberg:1994rs}.

In particular, Claus Montonen and David Olive argued in 1977 that the electric-magnetic duality (\ref{equ:duality}) 
should be valid not only in electrodynamics, but also in the SO(3) Georgi-Glashow model with 't~Hooft-Polyakov monopoles~\cite{Montonen:1977sn}. There arguments were based on the semiclassical approximation, and it is difficult to test their conjectured duality in the quantum theory, because of the problems associated with describing magnetic monopoles in quantum field theory~\cite{Rajantie:2005hi}. However, it has been shown to be true in supersymmetric versions of the theory~\cite{Seiberg:1994rs}, in which calculations are simpler.
This Montonen-Olive duality is very useful for many calculations, because it turns the strong magnetic coupling $\alpha_M$ into a weak electric coupling. Therefore one can use perturbation theory to do calculations even when interactions are strong.
Similar dualities have also turned out to be very important in superstring theory~\cite{Polchinski:1996nb}
where they can be used to relate different variants of the theory to each other. This suggests that there is actually only one fundamental theory, known as M-theory~\cite{Duff:1996aw}, and different superstring theories are simply different ways of describing the same physics.

\section{Conclusions}
One of the most fundamental properties of magnetism is that there are no magnetic charges, or magnetic monopoles. Yet, as we have seen, there are strong theoretical reasons to expect that they should exist.
Magnetic monopoles are considered to be ``one of the safest bets one can make about physics not yet seen''~\cite{Polchinski:2003bq}.
Theorists believe that they are simply so rare that we have not detected them yes, and so heavy that we have not been able to produce them in particle accelerators. 
However, as Petrus Peregrinus wrote~\cite{history2}, ``experience rather than argument is the basis of certainty in science'', and therefore the search has to continue. The discovery of magnetic monopoles would not only confirm that we are on the right track but also give us a unique way to probe physics at extremely high energies, far beyond the reach of any foreseeable particle accelerators.

Nevertheless, even without such a discovery, magnetic monopoles have been driving our understanding of physics towards more and more fundamental levels for centuries, and this progress is still continuing today in quantum field theory, where monopoles and the electric-magnetic duality give us access to phenomena we would not be able to study and understand in other ways.
The absence of magnetic monopoles in the universe also led to the theory of inflation, which is the cornerstone of modern cosmology.
Therefore, whether they exist or not, magnetic monopole have a central place in our understanding of the universe. 

\section*{Acknowledgements}
The author would like to thank David Weir for extensive discussions, and acknowledge funding by the STFC grant ST/J000353/1 and Royal Society International Joint Project JP100273. The calculations for Fig.~\ref{fig:lattice} were performed on the COSMOS Consortium supercomputer within the DiRAC Facility jointly funded by STFC and the Large Facilities Capital Fund of BIS.

\section*{Notes on contributor}
Arttu Rajantie is Reader in Theoretical Physics at Imperial College London. He obtained his PhD from the University of Helsinki in 1997. After that he worked for two years at the University of Sussex, and then for five years in the Department of Applied Mathematics and Theoretical Physics (DAMTP) at the University of Cambridge, where he was also a fellow of Churchill College, until he joined Imperial College in 2005.
He is interested in non-linear phenomena in quantum field theory, and their implications for particle physics and cosmology. In particular, he has developed ways to study magnetic monopoles and other topological solitons using lattice Monte Carlo simulations.

\bibliography{Monopoles_Rajantie}{}
\bibliographystyle{tCPH}
\label{lastpage}

\end{document}